\pdfoutput=1
\documentclass[journal,twoside,web]{ieeecolor}
\usepackage{tmi2}
\usepackage{cite}
\usepackage{amsmath,amssymb,amsfonts}
\usepackage{graphicx}
\usepackage{makecell}
\usepackage{multirow}
\usepackage[font=small,justification=justified,belowskip=2pt,aboveskip=2pt]{caption}

\def\baselinestretch{1.0}
\renewcommand{\arraystretch}{1.3}
\usepackage[table]{xcolor}

\usepackage[utf8]{inputenc}

\usepackage{hyperref}
\hypersetup{colorlinks=true,citecolor=teal,linkcolor=blue,urlcolor=blue}

\makeatletter
\def\widebreve{\mathpalette\wide@breve}
\def\wide@breve#1#2{\sbox\z@{$#1#2$}%
     \mathop{\vbox{\m@th\ialign{##\crcr
\kern0.08em\brevefill#1{0.8\wd\z@}\crcr\noalign{\nointerlineskip}%
                    $\hss#1#2\hss$\crcr}}}\limits}
\def\brevefill#1#2{$\m@th\sbox\tw@{$#1($}%
  \hss\resizebox{#2}{\wd\tw@}{\rotatebox[origin=c]{90}{\upshape(}}\hss$}

\providecommand{\slimits@}{\flatlimits}
\providecommand{\Cup}{\bigcup}

\newcommand*{\bigCup}{\mathop{\mathpalette\big@Cup\relax}\slimits@}
\newcommand*{\big@Cup}[2]{%
   \setbox\z@=\hbox{\m@th$#1\Cup$}%
   \setbox\z@=\vtop{\vbox{\kern.2\ht\z@\copy\z@}\kern.1\ht\z@}%
   \setbox\tw@=\hbox{\m@th$#1\bigcup$}%
   \vcenter{\hbox{\resizebox{!}{1.4\ht\tw@}{\box\z@}}}%
}
\makeatother

\definecolor{brightcerulean}{rgb}{0.11, 0.62, 0.74}

\def\BibTeX{{\rm B\kern-.05em{\sc i\kern-.025em b}\kern-.08em
    T\kern-.1667em\lower.7ex\hbox{E}\kern-.125emX}}

\begin{document}

\title{Stochastic Finite-Transition Learning and Exploration for Accelerated MRI Reconstruction}

\author{M. Usama Mirza, Esin O. Isik, and Tolga \c{C}ukur 
\vspace{-1.3cm}
\thanks{This study was supported in part by TUBA GEBIP 2015 and BAGEP 2017 fellowships, and by the TUBITAK 1004 Center of Excellence Support Program under Project No. 22AG016 awarded to E.O. Isik and T. \c{C}ukur (Corresponding author: Tolga \c{C}ukur, cukur@ee.bilkent.edu.tr).}
\thanks{M.U. Mirza and T. Cukur are with the Dept. of Electrical-Electronics Engineering and National Magnetic Resonance Research Center (UMRAM), Bilkent University, Ankara, Turkey, 06800. E.O. Isik is with the Institute of Biomedical Engineering, Bogazici University, Istanbul, Turkey, 34684.}
}

\maketitle

\begin{abstract}
Generative MRI reconstruction maps an initial source distribution to the target image distribution along a learned recovery trajectory. Common generative methods, such as diffusion and flow, typically learn fine-grained local dynamics whose integration defines this global mapping, hence requiring many network evaluations at inference. Existing few-step sampling approaches can lower computational burden by approximating large transformations from local dynamics, albeit at the expense of reconstruction fidelity. FLEX-MRI instead partitions recovery into a small number of finite intervals between retained states defined by progressive changes in k-space support, and directly learns the reconstruction change across each interval. Because such finite transitions can admit multiple plausible outcomes, FLEX-MRI combines support-relaxed finite-transition learning with stochastic exploration. During training, randomized k-space sampling masks define finite-transition targets with varying support, while the prediction loss remains unrestricted by each individual mask. At inference, FLEX-MRI perturbs a random subset of currently estimated k-space coefficients, allowing the learned predictor to stochastically explore alternative transition paths. Each predicted state is then reconciled with the acquired measurements through data consistency. We theoretically show that this stochastic exploration induces nonzero transition-level variation through the learned recovery map and data consistency without requiring an auxiliary generative sampler. Experiments on multi-coil MRI datasets show that FLEX-MRI outperforms competing methods while maintaining computational efficiency comparable to few-step baselines.
\end{abstract}
\vspace{-0.2cm}
\begin{IEEEkeywords} generative, MRI, reconstruction, stochastic, finite transition \vspace{-0.2cm}
\end{IEEEkeywords}

\bstctlcite{IEEEexample:BSTcontrol}

\section{Introduction}
Accelerated MRI reduces scan time by acquiring a subset of k-space coefficients \cite{lustig2007sparse}, leaving the unacquired coefficients to be recovered during reconstruction from the available measurements \cite{Zhang2020, fessler2020optimization}. Because undersampled measurements do not uniquely determine the missing information, multiple image estimates may remain compatible with the acquired data. This ambiguity has motivated the increasing use of generative priors \cite{narnhofer2019inverse, korkmaz2022unsupervised}, which constrain reconstruction using the distributional characteristics of high-quality MR images \cite{pixelcnnrecon, darestani2021accelerated, jalaln2021nips, zhao2024adaptive}. However, effective generative reconstruction depends not only on the learned image distribution, but also on how the recovery process itself is formulated from the acquired measurements toward the desired image. This formulation determines what intermediate recovery problems the model learns to solve and how they are traversed during inference, thereby affecting both reconstruction fidelity and computational cost.

Common generative methods, such as diffusion and flow, learn local recovery dynamics whose integration defines the global source-to-target mapping. Diffusion realizes this mapping over many intermediate states, allowing small recovery transitions and repeated enforcement of the acquired measurements \cite{song2022, chung2022media, cao2022highfrequency, luo2022mri}, but typically requiring hundreds of sequential network evaluations at inference \cite{peng2022towards, gungor2022adaptive}. MRI-specific formulations improve interaction with the acquired data through approaches including posterior guidance \cite{chungDPS}, explicit data consistency \cite{chung2023direct, ozturkler2023SMRD}, k-space-aware conditioning \cite{xie2022kspace, cao2022highfrequency, bian2024}, or task-aligned bridges between undersampled and fully-sampled states \cite{fabian2023diracdiffusion, mirza2026learning, gao2025selfdb}. While these formulations can improve conditioning across the recovery trajectory, they nevertheless retain the underlying fine-grained local recovery dynamics and their associated inference cost.

\begin{figure*}[t]
    \centering
    \begin{minipage}[t]{0.20\textwidth}
        \vspace{0pt}
        \captionsetup{justification=justified,singlelinecheck=false}
        \caption{Comparison of few-step recovery in \textbf{(a)} flow-based modeling, \textbf{(b)} transition matching, and \textbf{(c)} FLEX-MRI. Flow maps a fixed current state to a single next state through a learned velocity field, whereas transition matching models alternative outcomes using an auxiliary generative sampler with multiple inner steps per retained transition. FLEX-MRI instead perturbs the current state through random k-space undersampling and maps the resulting perturbed states with the same finite-transition predictor, enabling stochastic transition-level exploration without an auxiliary sampler or inner sampling loop.}
        \label{fig:flex_mri}
    \end{minipage}
    \hfill
    \begin{minipage}[t]{0.78\textwidth}
        \vspace{0pt}
        \centering
        \includegraphics[width=\linewidth]{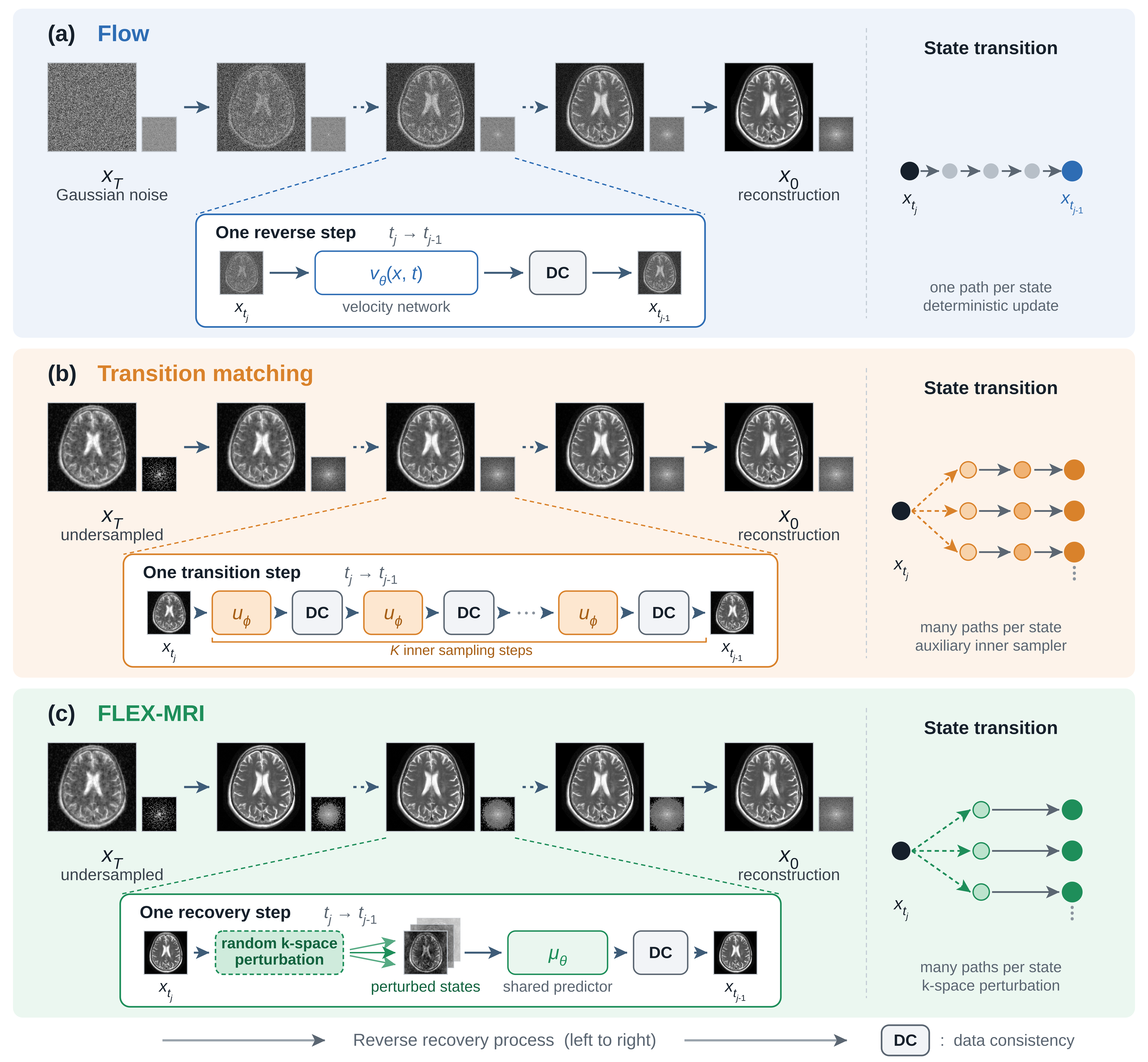}
    \end{minipage}
\end{figure*}

Existing few-step sampling approaches attempt to mitigate this burden by compressing fine-grained local recovery dynamics into coarser sampling steps, allowing the recovery trajectory to be traversed with fewer network evaluations. Diffusion acceleration realizes this through implicit sampling, distillation, or consistency-based formulations \cite{song2021ddim, salimans2022progressive, song2023consistency}, while flow-based models numerically integrate a continuous local velocity field using relatively few solver steps \cite{lipman2022flow, albergo2022building, liu2022flow, zhang2024flow, luo2025unsupervised}. In both cases, the transitions executed during few-step inference remain approximations derived from recovery dynamics learned at a finer scale, which can compromise reconstruction fidelity (Fig.~\ref{fig:flex_mri}a). The recent finite-transition framework instead changes the scale of the learned recovery problem by selecting a small set of retained states and directly learning the mapping across each retained interval. Transition matching follows this formulation by treating each finite transition as a conditional generative mapping \cite{shaul2025transition}. However, this mapping is implemented via an auxiliary generative sampler within each finite transition to represent transition-level variation, increasing the computational cost of each transition (Fig.~\ref{fig:flex_mri}b).

Here we introduce FLEX-MRI, a physics-guided finite-transition method for accelerated MRI reconstruction (Fig.~\ref{fig:flex_mri}c). FLEX-MRI defines retained states through progressive changes in k-space support and directly learns the finite transition across each retained interval. Thus, the finite transitions executed during few-step inference are themselves training targets. During training, randomized k-space sampling masks define finite-transition targets with varying k-space support, while support-relaxed finite-transition learning evaluates the prediction loss broadly across the image without restricting it to the coefficients selected by each individual mask. At inference, random k-space perturbation of the current reconstruction enables stochastic exploration of alternative transition outcomes before the acquired measurements are restored through data consistency. Our theoretical analysis shows that finite-transition recovery comprises a mean displacement together with residual variation among possible transition outcomes (Fig.~\ref{fig:finite_step}), and that stochastic exploration induces a non-degenerate residual within the degrees of freedom unresolved by the acquired measurements. FLEX-MRI thereby realizes transition-level variation through the finite-transition predictor itself, without requiring an auxiliary generative sampler. Code is available at {\small \href{https://github.com/icon-lab/FLEX-MRI}{https://github.com/icon-lab/FLEX-MRI}.}

\vspace{0.2cm}
\subsubsection*{\textbf{Contributions}}
\begin{itemize}

\item We introduce finite-transition modeling for accelerated MRI, where progressive changes in k-space support define retained states and the finite transition across each retained interval is learned directly. Support-relaxed finite-transition learning enables prediction across the image without restricting supervision to the coefficients selected by each individual transition mask.

\item We introduce stochastic exploration through random k-space perturbation of the current reconstruction before each finite-transition prediction, allowing the learned predictor to explore alternative transition outcomes without requiring an auxiliary generative sampler.

\item We theoretically show that stochastic exploration induces a non-degenerate residual through the same form of randomized k-space support change as the target finite-transition variation. The induced residual enters the same mean-plus-residual decomposition and occupies the same measurement-null subspace.

\end{itemize}

\section{Related Work}

\subsection{Many-Step Generative Recovery}

Diffusion models have become prominent generative priors for MRI reconstruction by defining fine-grained local recovery dynamics along a gradual transformation between a target distribution of high-quality MR images and a simple source distribution, typically white Gaussian noise \cite{song2020score,song2022,chung2022media}. Integration of these dynamics over many intermediate states defines the global source-to-target mapping, requiring repeated network evaluations during reconstruction. Acquired measurements can be incorporated throughout this trajectory through data-consistency projections and/or acquisition-aware model conditioning \cite{chungDPS,chung2023direct,xie2022kspace}. The fine discretization provides repeated opportunities to refine uncertain image estimates and reconcile them with the acquired measurements, but requires a large number of sequential network evaluations.

Recent diffusion formulations have also considered task-aligned source and target distributions connected through intermediate states derived from the reconstruction problem rather than a generic noise-to-image process \cite{liu2023i2sb,huang2023cdiffmr}. In accelerated MRI, such trajectories can directly connect image states associated with undersampled and fully-sampled acquisitions through progressive changes in k-space sampling \cite{mirza2026learning}. This task alignment places the recovery trajectory closer to the underlying acquisition process, but retains the fine-grained local recovery dynamics and therefore does not necessarily reduce the number of transitions required for recovery.

\subsection{Few-Step Generative Recovery}
Diffusion acceleration methods reduce the number of network evaluations while retaining a recovery formulation based on fine-grained local recovery dynamics. Implicit samplers traverse the trajectory learned from finely-spaced image states using fewer retained states \cite{song2021ddim}, distillation transfers the cumulative effect of multiple fine transitions from a more finely discretized teacher \cite{salimans2022progressive}, and consistency-based formulations encourage predictions from distant states to agree with a common recovery target \cite{song2023consistency}. These approaches therefore reduce sampling cost by approximating finite transitions from recovery dynamics originally learned at a finer scale. As the spacing between retained states increases, larger discretization steps can make the Gaussian reverse-kernel approximation increasingly inaccurate \cite{guo2023gaussian}, contributing to fidelity loss under aggressive sampling acceleration.

Flow models face the same mismatch between the scale of learned recovery dynamics and few-step inference. They learn a continuous local velocity field that specifies the transport direction at each state \cite{lipman2022flow,liu2022flow,albergo2022building}, while numerical integration with relatively few solver steps approximates finite transitions from this locally defined field \cite{zhang2024flow}. As the solver steps become larger, the transformations executed during inference increasingly span intervals that were not learned directly, which can compromise reconstruction fidelity. Flow models additionally provide a single transport direction for a given current state (Fig.~\ref{fig:flex_mri}a), so integration commonly follows a deterministic recovery path \cite{lipman2022flow}, potentially causing errors introduced by an earlier transition persist across subsequent steps.

\subsection{Finite-Transition Modeling}
Diverging from common generative methods, finite-transition modeling changes the scale at which the recovery problem is learned. Rather than deriving finite transitions from recovery dynamics learned at a finer scale, it directly models the conditional mapping across each retained interval \cite{shaul2025transition}. To represent transition-level variation between widely separated retained states, however, it introduces an auxiliary generative sampler within each finite transition (Fig.~\ref{fig:flex_mri}b), partially offsetting the efficiency gained from using fewer states. Common transition-matching formulations also define intermediate states through linear interpolation between source and target endpoints \cite{shaul2025transition}. Such interpolation is mismatched to accelerated MRI, where k-space acquisition is inherently binary and intermediate undersampling patterns cannot be represented by continuously averaging undersampled and fully-sampled data.

FLEX-MRI adopts finite-transition modeling but defines its retained states and transition-level variation directly from the MRI acquisition process (Fig.~\ref{fig:flex_mri}c). Progressive changes in k-space support define the retained states, while the finite transition across each retained interval is directly supervised during training. Randomized k-space sampling masks provide varying finite-transition targets, and support-relaxed finite-transition learning evaluates the prediction loss without restriction to the coefficients selected by each individual mask. At inference, random k-space perturbation enables stochastic exploration of alternative transition outcomes before data consistency restores agreement with the acquired measurements. FLEX-MRI thereby combines direct finite-transition learning with stochastic exploration without an auxiliary generative sampler.

\section{Theory}
The theoretical analysis of FLEX-MRI is organized in two parts. We first formulate few-step MRI recovery from a finite-transition perspective and show that omitting intermediate states introduces transition-level variation that must be represented within each retained transition. We then realize this stochastic transition behavior without auxiliary generative sampling, and establish the correspondence between the variation induced by stochastic k-space perturbation and that required by the finite-transition process. The key notation used is described in Table \ref{tab:key_notation}.

\begin{table}[t]
\centering
\caption{Key notation for FLEX-MRI.}
\label{tab:key_notation}
\begin{tabular}{p{0.18\linewidth} p{0.74\linewidth}}
\hline
\textbf{Symbol} & \textbf{Definition} \\
\hline

$\mathbf{x}_{t_j}(\boldsymbol{\rho})$
& Forward-process state at retained index $t_j$. \\

$\hat{\mathbf{x}}_{t_j}$
& Current reconstruction state at $t_j$. \\

$\tilde{\mathbf{x}}_{t_j}^{(z)}$
& Perturbed current state obtained by additional stochastic undersampling
across k-space. \\

$\boldsymbol{\Delta}_{t_{j-1}\leftarrow t_j}$
& Finite recovery increment from retained state $t_j$ to $t_{j-1}$. \\

$\mathbf{d}_{t_{j-1}\leftarrow t_j}$
& Mean displacement of the retained finite transition. \\

$\mathbf{r}_{t_{j-1}\leftarrow t_j}$
& Residual variation of the retained finite transition. \\

$\boldsymbol{\eta}_j$
& Binary transition mask identifying the unacquired k-space coefficients
that change in retained transition $j$. \\

$\mu_\theta(\cdot,t_{j-1})$
& Finite-transition predictor from $t_j$ to $t_{j-1}$. \\

$\mathcal{P}_{\gamma_j}(\cdot;\mathbf{z}_j)$
& Stochastic perturbation operator for transition $j$. \\

$\gamma_j$
& Perturbation rate for stochastic exploration at transition $j$. \\

$\boldsymbol{\Pi}_{\mathbf{y}}$
& Data-consistency projection enforcing agreement with the acquired
measurements after each predicted transition. \\

\hline
\end{tabular}
\end{table}

\subsection{Finite-Transition View of Few-Step MRI Recovery}
\label{subsec:finite-transition-theory}

Recent task-aligned generative MRI priors formulate recovery between undersampled $\mathbf{x}_T$ and fully sampled $\mathbf{x}_0$ endpoints through a forward supervision process across a set of intermediate states \cite{fabian2023diracdiffusion, mirza2026learning}, $\mathbf{x}_0 \rightarrow \mathbf{x}_1 \rightarrow \cdots
\rightarrow \mathbf{x}_T$. A network-based predictor is then learned to implement the corresponding reverse process, $\mathbf{x}_T \rightarrow \mathbf{x}_{T-1} \rightarrow \cdots \rightarrow \mathbf{x}_0$. For a finely discretized supervision process ($T\gg1$), consecutive states are relatively similar and each network evaluation therefore only has to estimate a fine-transition (Fig. \ref{fig:finite_step}a). However, inference typically requires hundreds to thousands of sequential evaluations that are computationally burdening.

To improve computational efficiency, few-step recovery instead retains only a subset of the original states,
\begin{equation}
T=t_J>t_{J-1}>\cdots>t_1>t_0=0,
\qquad
1\leq J\ll T,
\label{eq:retained_states}
\end{equation}
where each $t_j$ is an index on the original fine-grained trajectory (Fig. \ref{fig:finite_step}b). Let $\mathbf{y}=\mathbf{A}\mathbf{x}_0$ denote the measured undersampled k-space data. For clarity of the mask analysis below, we suppress coil-sensitivity encoding and use $\Omega$ to denote the Fourier coordinates sampled by $\mathbf{A}$; the multi-coil implementation applies the corresponding encoding and data-consistency operations. For a retained transition $t_j\rightarrow t_{j-1}$, the finite recovery increment can be defined as:
\begin{equation}
\boldsymbol{\Delta}_{t_{j-1}\leftarrow t_j}
\triangleq
\mathbf{x}_{t_{j-1}}-\mathbf{x}_{t_j},
\label{eq:finite_increment}
\end{equation}
where recovery is governed by the conditional distribution
$q_{t_{j-1}\leftarrow t_j}
(\mathbf{x}_{t_{j-1}}\mid\mathbf{x}_{t_j},\mathbf{y})$ (we neglect $\mathbf{y}$ in the equations below for brevity). This increment can be decomposed into mean displacement and residual variation terms as  (Fig. \ref{fig:finite_step}c):
\begin{align}
\boldsymbol{\Delta}_{t_{j-1}\leftarrow t_j}
&=
\mathbf{d}_{t_{j-1}\leftarrow t_j}(\mathbf{x}_{t_j})
+
\mathbf{r}_{t_{j-1}\leftarrow t_j}; \nonumber \\
\label{eq:finite_decomposition}
&\mathbf{d}_{t_{j-1}\leftarrow t_j}(\mathbf{x}_{t_j})
\triangleq
\mathbb{E}
\left[
\boldsymbol{\Delta}_{t_{j-1}\leftarrow t_j}
\mid
\mathbf{x}_{t_j}
\right], \nonumber \\
&\mathbf{r}_{t_{j-1}\leftarrow t_j}
\triangleq
\boldsymbol{\Delta}_{t_{j-1}\leftarrow t_j}
-
\mathbf{d}_{t_{j-1}\leftarrow t_j}(\mathbf{x}_{t_j}).
\end{align}
As retained states preserve the acquired coefficients in $\Omega$, $\mathbf{A}\boldsymbol{\Delta}_{t_{j-1}\leftarrow t_j}=\mathbf{0}$ for every retained transition. Linearity of conditional expectation then gives $\mathbf{A}\mathbf{r}_{t_{j-1}\leftarrow t_j}=\mathbf{0}$, so the residual variation lies within the measurement-null subspace.

The residual term can be related directly to the omitted intermediate states. Consider a fine-grid state $u$ satisfying $t_{j-1}<u<t_j$. Under a Markov formulation, the retained transition marginalizes over the possible intermediate states \cite{holderrieth2025generator, shaul2025transition},
\begin{align}
q_{t_{j-1}\leftarrow t_j}
(\mathbf{x}_{t_{j-1}}\mid\mathbf{x}_{t_j})
&=
\int
q_{t_{j-1}\leftarrow u}
(\mathbf{x}_{t_{j-1}}\mid\mathbf{x}_u)
\nonumber\\
&\qquad\qquad
q_{u\leftarrow t_j}
(\mathbf{x}_u\mid\mathbf{x}_{t_j})
\,d\mathbf{x}_u .
\label{eq:finite_transition_composition}
\end{align}
For a fixed $\mathbf{x}_{t_j}$, different possible values of $\mathbf{x}_u$ can therefore lead to different recovery outcomes. To make this effect explicit, define the covariance of the residual term as
\begin{equation}
\boldsymbol{\Sigma}_{t_{j-1}\leftarrow t_j}(\mathbf{x}_{t_j})
\triangleq
\operatorname{Cov}
\left[
\mathbf{r}_{t_{j-1}\leftarrow t_j}
\mid
\mathbf{x}_{t_j}
\right].
\label{eq:finite_covariance}
\end{equation}
Applying the law of total covariance with respect to $\mathbf{x}_{u}$ gives
\begin{align}
\boldsymbol{\Sigma}_{t_{j-1}\leftarrow t_j}(\mathbf{x}_{t_j})
&=
\mathbb{E}_{\mathbf{x}_u\mid\mathbf{x}_{t_j}}
\left[
\boldsymbol{\Sigma}_{t_{j-1}\leftarrow u}(\mathbf{x}_u)
\right]
\nonumber\\
&\quad+
\operatorname{Cov}_{\mathbf{x}_u\mid\mathbf{x}_{t_j}}
\left[
\mathbf{x}_u+
\mathbf{d}_{t_{j-1}\leftarrow u}(\mathbf{x}_u)
\right].
\label{eq:cov_composition}
\end{align}
The first term contains the residual variation of the finer transition
$u\rightarrow t_{j-1}$, while the second captures variation across the possible omitted states and their corresponding mean displacement. Hence, variation that would originally be represented across several smaller steps must be represented within the residual component of the retained transition \cite{guo2023gaussian}. 

Flow methods commonly map a fixed current state to a single next state through a learned continuous velocity field \cite{lipman2022flow, liu2022flow, albergo2022building} and hence do not explicitly represent a nonzero conditional residual term $\mathbf{r}_{t_{j-1}\leftarrow t_j}$. While transition matching represents such conditional variation through an auxiliary generative model \cite{shaul2025transition}, this introduces an additional inner sampling process, offsetting the computational benefit sought through few-step recovery. Effective few-step methods are therefore needed that preserve stochastic transition-level variation without elevating computational burden.

\subsection{FLEX-MRI}
\label{subsec:flex-theory}

\begin{figure*}[t]
    \centering
    \begin{minipage}[t]{0.22\textwidth}
        \vspace{0pt}
        \captionsetup{justification=justified,singlelinecheck=false}
        \caption{Finite-transition view of few-step recovery in FLEX-MRI. \textbf{(a)} Fine-step recovery traverses many small transitions with narrow conditional distributions. \textbf{(b)} Few-step recovery retains instead $J$$\ll$$T$ transitions with broader conditional distributions. \textbf{(c)} Marginalizing the omitted states yields a finite-transition increment comprising a mean displacement $\mathbf{d}$ and residual variation $\mathbf{r}$, whereas a deterministic transition captures only the mean. \textbf{(d)} FLEX-MRI realizes the residual variation through random k-space perturbation, a shared predictor $\mu_\theta(\cdot,t_{j-1})$, and data consistency. Multiple mask realizations illustrate the distribution of possible next states; one realization is used per transition at inference.}
    \label{fig:finite_step}
    \end{minipage}
    \hfill
    \begin{minipage}[t]{0.76\textwidth}
        \vspace{0pt}
        \centering
        \includegraphics[width=\linewidth]{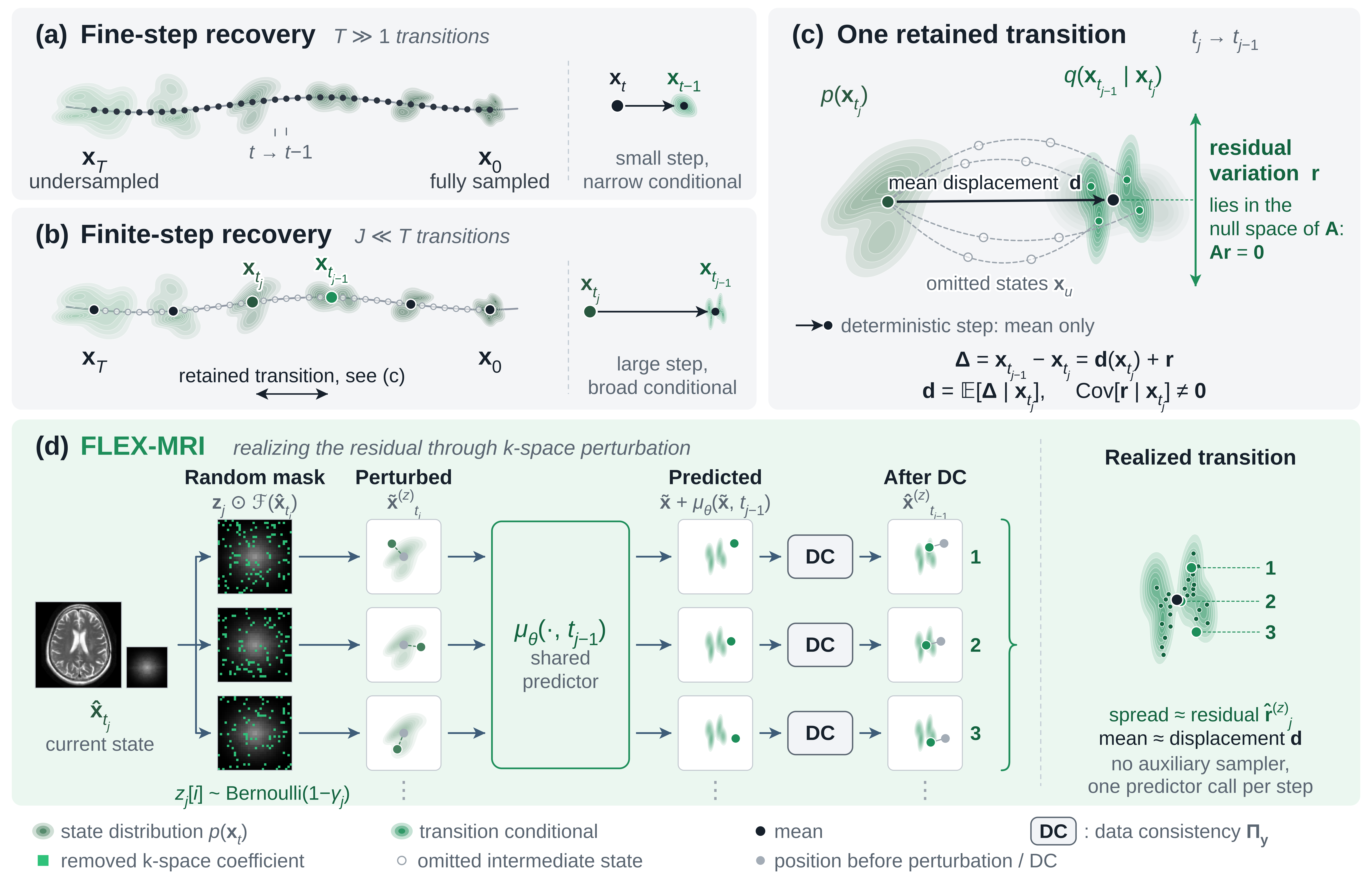}
    \end{minipage}
\end{figure*}

To realize this transition-level variation without auxiliary generative sampling, FLEX-MRI combines support-relaxed finite-transition learning with stochastic exploration  (Fig. \ref{fig:finite_step}d). During training, randomized k-space sampling masks with progressively growing undersampling rates define finite-transition targets, while the prediction loss is evaluated over the complete update rather than being restricted to the coefficients selected by each mask (Sec.~\ref{subsubsec:task_aligned_ft}). At inference, stochastic undersampling across k-space produces alternative inputs to the same finite-transition predictor (Sec.~\ref{subsubsec:stochastic_cross_transition}). We then characterize how the resulting conditional variation propagates through the predictor and data consistency (Sec.~\ref{subsubsec:conditional_variation}).

\subsubsection{Support-Relaxed Finite-Transition Learning}
\label{subsubsec:task_aligned_ft}

Prior work has reported benefits of diffusion processes based on progressive k-space removal for MRI reconstruction \cite{mirza2026learning}. Inspired by this finding, we construct the forward process of FLEX-MRI directly from progressive k-space undersampling and use the resulting intermediate states to define finite transitions. Let $\Omega$ denote the k-space locations available at the undersampled endpoint $\mathbf{x}_T$ and $\bar{\Omega}$ their complement. For each $i\in\bar{\Omega}$, an independent priority value $\rho_i\sim\mathcal{U}(0,1)$ is drawn. With $\nu_t$ denoting the removal threshold, the sampling mask is
\begin{equation}
m_t[i]
=
\begin{cases}
1, & i\in\Omega,\\
\mathbb{I}(\rho_i\geq\nu_t), & i\in\bar{\Omega},
\end{cases}
\label{eq:forward_mask}
\end{equation}
where $\nu_t$ increases along the forward trajectory, with $\nu_0=0$ and $\nu_T=1$. The corresponding image state is
\begin{equation}
\mathbf{x}_t(\boldsymbol{\rho})
=
\mathcal{F}^{-1}
\left[
\mathbf{m}_t(\boldsymbol{\rho})
\odot
\mathcal{F}(\mathbf{x}_0)
\right].
\label{eq:forward_state_random}
\end{equation}
For each forward state, $\mathbf{m}_t$ specifies which k-space coefficients remain sampled at that point along the trajectory. A finite transition $t_j\rightarrow t_{j-1}$ is therefore determined by the change between the sampling masks of its associated states. We define the corresponding transition mask as
\begin{equation}
\eta_j[i]
\triangleq
m_{t_{j-1}}[i]-m_{t_j}[i]
=
\begin{cases}
0, & i\in\Omega,\\
\mathbb{I}
\left(
\nu_{t_{j-1}}
\leq
\rho_i
<
\nu_{t_j}
\right), & i\in\bar{\Omega}.
\end{cases}
\label{eq:forward_transition_mask}
\end{equation}
With $a_j\triangleq\nu_{t_j}-\nu_{t_{j-1}}$, each unacquired coefficient in the transition mask satisfies
\begin{equation}
\eta_j[i]
\sim
\operatorname{Bernoulli}(a_j),
\qquad
i\in\bar{\Omega}.
\label{eq:forward_transition_bernoulli}
\end{equation}
The removal thresholds are equispaced along the forward trajectory, so that a fixed number of randomly selected coefficients is removed at each step.

The finite-transition target generated by a particular realization of the forward process is therefore
\begin{align}
\boldsymbol{\Delta}^{0}_j(\boldsymbol{\rho}) \triangleq \mathbf{x}_{t_{j-1}}(\boldsymbol{\rho}) - \mathbf{x}_{t_j}(\boldsymbol{\rho}) = \mathcal{F}^{-1}
\left[ \boldsymbol{\eta}_j(\boldsymbol{\rho}) \odot \mathcal{F}(\mathbf{x}_0) \right].
\label{eq:forward_transition_increment}
\end{align}
For a given realization, $\boldsymbol{\eta}_j$ identifies the coefficient changes that define the transition from $\mathbf{x}_{t_j}$ to $\mathbf{x}_{t_{j-1}}$. Following the theory of discrete conditional transitions in generative modeling \cite{shaul2025transition}, the restricted support of Fourier representations \cite{yaman2020, yaman2022multimask} implies the following regression objective that compares the predicted and target changes over k-space coefficients:
\begin{align}
\mathcal{L}_{\mathrm{FT}}
&=
\mathbb{E}_{\mathbf{x}_0,\boldsymbol{\rho},j}
\bigg[
\Big\|
\boldsymbol{\eta}_j(\boldsymbol{\rho})
\odot
\Big(
\mathcal{F}
\left[
\mu_\theta
\left(
\mathbf{x}_{t_j},t_{j-1}
\right)
\right]
-
\mathcal{F}
\left[
\boldsymbol{\Delta}^{0}_j
\right]
\Big)
\Big\|_2^2
\bigg].
\label{eq:restricted_transition_objective}
\end{align}
Note that the transition mask controls both the definition of the transition target and the coefficients over which the corresponding prediction error is evaluated.

To enable support-relaxed learning, FLEX-MRI relaxes the transition-specific restriction in Eq. \ref{eq:restricted_transition_objective} by removing the transition mask from the prediction error. Accordingly, the unrestricted Fourier-domain objective becomes
\begin{align}
\mathcal{L}_{\mathrm{FLEX}} &= \mathbb{E}_{\mathbf{x}_0,\boldsymbol{\rho},j}
\left[\left\|\mathcal{F}\left[\mu_\theta\left(\mathbf{x}_{t_j},t_{j-1}\right)\right]-\mathcal{F}\left[\boldsymbol{\Delta}^{0}_j\right]\right\|_2^2\right], \nonumber \\
&= \mathbb{E}_{\mathbf{x}_0,\boldsymbol{\rho},j}\left[ \left\| \mu_\theta \left( \mathbf{x}_{t_j},t_{j-1} \right) - \boldsymbol{\Delta}^{0}_j \right\|_2^2 \right].
\label{eq:finite_transition_objective}
\end{align}
where the image-domain equivalent in the second line follows from Parseval's theorem for a unitary Fourier transform. Because $\boldsymbol{\Delta}^{0}_j$ is generated from a different randomized transition mask for each realization, the predictor does not associate a retained transition index with a fixed subset of unacquired coefficients. Specifically,
\begin{equation}
\mathbb{E}_{\boldsymbol{\rho}}\left[ \boldsymbol{\eta}_j \right] = a_j\mathbf{m}_{\bar{\Omega}}, \qquad \Pr_{\boldsymbol{\rho}} \left( \eta_j[i]=1 \right) = a_j>0, \quad
i\in\bar{\Omega}.
\label{eq:transition_mask_coverage}
\end{equation}
Across $L$ independent realizations of the forward process, the probability that coefficient $i$ is selected by the transition mask for transition $j$ at least once is
\begin{equation}
\Pr \left( \sum_{\ell=1}^{L} \eta_j^{(\ell)}[i] > 0 \right) = 1-(1-a_j)^L, 
\label{eq:transition_mask_union}
\end{equation}
approaching unity as $L$ increases. \textit{Each transition is therefore learned from targets generated by different randomized transition masks spanning the unacquired k-space, while support-relaxed learning removes the corresponding restriction from the prediction loss.} Variation among these finite-transition targets is generated by stochastic changes in k-space support through $\boldsymbol{\eta}_j$ and is confined to $\bar{\Omega}$, the same measurement-null degrees of freedom in which $\mathbf{r}_{t_{j-1}\leftarrow t_j}$ varies.

\subsubsection{Stochastic Finite-Transition Recovery}
\label{subsubsec:stochastic_cross_transition}

Support-relaxed finite-transition learning removes the transition-mask restriction from the objective, but direct application of the learned predictor at inference remains deterministic. For a fixed current state, $\mu_\theta$ produces a single transition, so recovery from the fixed endpoint $\mathbf{x}_T$ follows the same sequence of predicted states and cannot realize the residual variation $\mathbf{r}_{t_{j-1}\leftarrow t_j}$ \cite{shaul2025transition}. FLEX-MRI introduces stochasticity into this recovery by perturbing the current state through additional k-space undersampling prior to each transition.

Given the current state $\hat{\mathbf{x}}_{t_j}$, we randomly undersample its k-space coefficients to form
\begin{equation}
\tilde{\mathbf{x}}_{t_j}^{(z)}
=
\mathcal{P}_{\gamma_j}
\left(
\hat{\mathbf{x}}_{t_j};
\mathbf{z}_j
\right)
\triangleq
\mathcal{F}^{-1}
\left[
\mathbf{z}_j
\odot
\mathcal{F}
\left(
\hat{\mathbf{x}}_{t_j}
\right)
\right],
\label{eq:stochastic_perturb}
\end{equation}
where
\begin{equation}
z_j[i]
\sim
\operatorname{Bernoulli}(1-\gamma_j),
\qquad
i\in\Omega\cup\bar{\Omega},
\label{eq:stochastic_mask}
\end{equation}
with $\gamma_j$ denoting the perturbation rate. We choose this perturbation as it follows the same form of k-space degradation used to train the finite-transition predictor, while applying it across the complete k-space estimate at inference. Thus, training-target variation and inference-time stochastic exploration are both generated through randomized changes in k-space support. Unlike the masks in Eq.~\eqref{eq:forward_mask}, the inference-time perturbation can also remove coefficients in $\Omega$, which are subsequently restored by the DC projection in Eq.~\eqref{eq:flex_transition}.

Because the inference-time perturbation is sampled across all k-space locations, the perturbed coefficients need not coincide with those selected by any particular training transition mask. The learned predictor in Eq.~\eqref{eq:finite_transition_objective} can therefore operate on these perturbed states without requiring a matching transition mask. The provisional next state is then
\begin{equation}
\hat{\mathbf{x}}_{t_{j-1}}^{(z),\mathrm{pred}}
=
\tilde{\mathbf{x}}_{t_j}^{(z)}
+
\mu_\theta
\left(
\tilde{\mathbf{x}}_{t_j}^{(z)},t_{j-1}
\right).
\label{eq:stochastic_prediction}
\end{equation}
Afterwards, the acquired measurements are enforced through a data-consistency projection \cite{schlemper2017deep, aggarwal2018modl}:
\begin{equation}
\hat{\mathbf{x}}_{t_{j-1}}^{(z)}
=
\boldsymbol{\Pi}_{\mathbf{y}}
\left(
\hat{\mathbf{x}}_{t_{j-1}}^{(z),\mathrm{pred}}
\right),\,\,\,
\boldsymbol{\Pi}_{\mathbf{y}}(\mathbf{v})
=
\mathbf{v}
+
\mathbf{A}^{\dagger}
\left(
\mathbf{y}
-
\mathbf{A}\mathbf{v}
\right).
\label{eq:flex_transition}
\end{equation}

The overall mapping during a transition in FLEX-MRI can therefore be described through a transition increment $\hat{\boldsymbol{\Delta}}_j^{(z)}$ and its conditional residual $\hat{\mathbf{r}}_j^{(z)}$ as
\begin{align}
\hat{\boldsymbol{\Delta}}_j^{(z)}
\triangleq
\hat{\mathbf{x}}_{t_{j-1}}^{(z)}
-
\hat{\mathbf{x}}_{t_j}; \,\,\,
\hat{\mathbf{r}}_j^{(z)}
\triangleq
\hat{\boldsymbol{\Delta}}_j^{(z)}
-
\mathbb{E}_{\mathbf{z}_j}
\left[
\hat{\boldsymbol{\Delta}}_j^{(z)}
\mid
\hat{\mathbf{x}}_{t_j}
\right].
\label{eq:flex_residual_decomposition}
\end{align}
The FLEX-MRI transition takes the same mean-plus-residual form as Eq.~\eqref{eq:finite_decomposition}. Because data consistency enforces the same acquired measurements at $\hat{\mathbf{x}}_{t_j}$ and $\hat{\mathbf{x}}_{t_{j-1}}^{(z)}$, $\mathbf{A}\hat{\boldsymbol{\Delta}}_j^{(z)}=\mathbf{0}$ and therefore $\mathbf{A}\hat{\mathbf{r}}_j^{(z)}=\mathbf{0}$. \textit{Thus, the FLEX-MRI residual and $\mathbf{r}_{t_{j-1}\leftarrow t_j}$ in Eq.~\eqref{eq:finite_decomposition} share the same mean-plus-residual role and measurement-null support, with both variations arising from randomized changes in k-space support.} Fig.~\ref{fig:finite_step}(d) illustrates this process for several realizations of the perturbation mask applied to the same current state.

\subsubsection{Propagation of Conditional Variation}
\label{subsubsec:conditional_variation}

The remaining question is whether the FLEX-MRI residual retains nonzero variation after propagation through the predictor and data consistency. Conditioned on the current state $\hat{\mathbf{x}}_{t_j}$, the only randomness for the perturbed state in Eq.~\eqref{eq:stochastic_perturb} arises from $\mathbf{z}_j$. Since
$\mathbb{E}[z_j[i]]=1-\gamma_j$ and $\operatorname{Var}[z_j[i]]=\gamma_j(1-\gamma_j)$, the perturbed state can be expressed in Fourier domain as
\begin{align}
\mathcal{F}
\left(
\tilde{\mathbf{x}}_{t_j}^{(z)}
\right)
=& \,\,
\mathbb{E}_{\mathbf{z}_j}
\left[
\mathcal{F}
\left(
\tilde{\mathbf{x}}_{t_j}^{(z)}
\right)
\mid
\hat{\mathbf{x}}_{t_j}
\right]
\nonumber \\ &+
\left(
\mathbf{z}_j-(1-\gamma_j)\mathbf{1}
\right)
\odot
\mathcal{F}
\left(
\hat{\mathbf{x}}_{t_j}
\right).
\label{eq:centered_perturbed_kspace}
\end{align}
Independence of the Bernoulli variables then gives
\begin{align}
\operatorname{Cov}_{\mathbf{z}_j}
\left[
\mathcal{F}
\left(
\tilde{\mathbf{x}}_{t_j}^{(z)}
\right)
\mid
\hat{\mathbf{x}}_{t_j}
\right]
&=
\gamma_j(1-\gamma_j)
\operatorname{Diag}
\left(
\left|
\mathcal{F}
\left(
\hat{\mathbf{x}}_{t_j}
\right)
\right|^2
\right).
\label{eq:perturbed_kspace_covariance}
\end{align}
The corresponding image-domain covariance is obtained by applying the inverse Fourier transform to this k-space variation; we retain the compact covariance notation below.

To track this variation through the transition, we first locally linearize Eq.~\eqref{eq:stochastic_prediction} around $\hat{\mathbf{x}}_{t_j}$:
\begin{align}
\hat{\mathbf{x}}_{t_{j-1}}^{(z),\mathrm{pred}}
\approx\,&
\hat{\mathbf{x}}_{t_j}
+
\mu_\theta
\left(
\hat{\mathbf{x}}_{t_j},t_{j-1}
\right)
+
\mathbf{M}_j
\left(
\tilde{\mathbf{x}}_{t_j}^{(z)}
-
\hat{\mathbf{x}}_{t_j}
\right),
\label{eq:local_transition_expansion}
\end{align}
where
\begin{equation}
\mathbf{M}_j
\triangleq
\mathbf{I}
+
\left.
\frac{
\partial
\mu_\theta
\left(
\mathbf{x},t_{j-1}
\right)}
{\partial\mathbf{x}}
\right|_{\mathbf{x}=\hat{\mathbf{x}}_{t_j}} .
\label{eq:transition_jacobian}
\end{equation}
Subtracting the conditional mean of Eq.~\eqref{eq:local_transition_expansion} therefore gives
\begin{align}
&\hat{\mathbf{x}}_{t_{j-1}}^{(z),\mathrm{pred}}
-
\mathbb{E}_{\mathbf{z}_j}
\left[
\hat{\mathbf{x}}_{t_{j-1}}^{(z),\mathrm{pred}}
\mid
\hat{\mathbf{x}}_{t_j}
\right]
\nonumber\\
&\qquad\approx
\mathbf{M}_j
\left(
\tilde{\mathbf{x}}_{t_j}^{(z)}
-
\mathbb{E}_{\mathbf{z}_j}
\left[
\tilde{\mathbf{x}}_{t_j}^{(z)}
\mid
\hat{\mathbf{x}}_{t_j}
\right]
\right).
\label{eq:centered_transition_prediction}
\end{align}
The data-consistency projection subsequently acts upon this variation, with the difference between two provisional estimates based on Eq.~\eqref{eq:flex_transition} given as
\begin{align}
\boldsymbol{\Pi}_{\mathbf{y}}
\left(
\mathbf{v}^{(1)}
\right)
-
\boldsymbol{\Pi}_{\mathbf{y}}
\left(
\mathbf{v}^{(2)}
\right)
&=
\left(
\mathbf{I}
-
\mathbf{A}^{\dagger}\mathbf{A}
\right)
\left(
\mathbf{v}^{(1)}
-
\mathbf{v}^{(2)}
\right)
\nonumber\\
&\triangleq
\mathbf{P}_{\mathcal{N}}
\left(
\mathbf{v}^{(1)}
-
\mathbf{v}^{(2)}
\right).
\label{eq:null_projection}
\end{align}
Combining Eqs.~\eqref{eq:centered_transition_prediction} and \eqref{eq:null_projection}, the conditional covariance after the complete finite-transition step can be approximated as
\begin{align}
\operatorname{Cov}_{\mathbf{z}_j}
\left[
\hat{\mathbf{x}}_{t_{j-1}}^{(z)}
\mid
\hat{\mathbf{x}}_{t_j}
\right]
&\approx
\mathbf{P}_{\mathcal{N}}
\mathbf{M}_j
\operatorname{Cov}_{\mathbf{z}_j}
\left[
\tilde{\mathbf{x}}_{t_j}^{(z)}
\mid
\hat{\mathbf{x}}_{t_j}
\right]
\mathbf{M}_j^H
\mathbf{P}_{\mathcal{N}}^H .
\label{eq:flex_transition_covariance}
\end{align}
As $\hat{\mathbf{x}}_{t_j}$ is fixed under the conditional covariance, Eq.~\eqref{eq:flex_residual_decomposition} gives
\begin{equation}
\operatorname{Cov}_{\mathbf{z}_j}
\left[
\hat{\mathbf{r}}_j^{(z)}
\mid
\hat{\mathbf{x}}_{t_j}
\right]
=
\operatorname{Cov}_{\mathbf{z}_j}
\left[
\hat{\mathbf{x}}_{t_{j-1}}^{(z)}
\mid
\hat{\mathbf{x}}_{t_j}
\right].
\label{eq:flex_residual_covariance}
\end{equation}
Hence, Eq.~\eqref{eq:flex_transition_covariance} directly characterizes the conditional covariance of the FLEX-MRI residual.

The stochastic perturbation in Eq.~\eqref{eq:stochastic_perturb} generally contains components associated with both acquired and unacquired k-space coefficients. Accordingly, let the centered perturbation $\boldsymbol{\delta}_j$ be decomposed as $\boldsymbol{\delta}_j=\boldsymbol{\delta}_{j,\bar{\Omega}}+\boldsymbol{\delta}_{j,\Omega}$, such that $\boldsymbol{\delta}_{j,\bar{\Omega}}=\mathbf{P}_{\mathcal{N}}\boldsymbol{\delta}_j$, $\boldsymbol{\delta}_{j,\Omega}=\left(\mathbf{I}-\mathbf{P}_{\mathcal{N}}\right)\boldsymbol{\delta}_j$.
Using Eq.~\eqref{eq:transition_jacobian}, propagation through the predictor and data consistency gives
\begin{align}
\mathbf{P}_{\mathcal{N}}\mathbf{M}_j\boldsymbol{\delta}_j = \left( \mathbf{I} + \mathbf{P}_{\mathcal{N}}\mathbf{J}_j \right) \boldsymbol{\delta}_{j,\bar{\Omega}}
+\mathbf{P}_{\mathcal{N}}\mathbf{J}_j
\boldsymbol{\delta}_{j,\Omega},
\label{eq:variation_propagation}
\end{align}
where $\mathbf{J}_j\triangleq\mathbf{M}_j-\mathbf{I}$ denotes the predictor Jacobian appearing in Eq.~\eqref{eq:transition_jacobian}. Because the components associated with $\Omega$ and $\bar{\Omega}$ are sampled independently across disjoint sets of k-space locations, they have zero cross-covariance. Denoting the respective covariance matrices by $\boldsymbol{\Sigma}_{\Omega}$ and $\boldsymbol{\Sigma}_{\bar{\Omega}}$, Eq.~\eqref{eq:flex_transition_covariance} becomes
\begin{align}
\boldsymbol{\Sigma}_{\mathrm{out}}
=
\left(
\mathbf{I}
+
\mathbf{P}_{\mathcal{N}}\mathbf{J}_j
\right)
\boldsymbol{\Sigma}_{\bar{\Omega}}
\left(
\mathbf{I}
+
\mathbf{P}_{\mathcal{N}}\mathbf{J}_j
\right)^H
+
\mathbf{P}_{\mathcal{N}}\mathbf{J}_j
\boldsymbol{\Sigma}_{\Omega}
\mathbf{J}_j^H
\mathbf{P}_{\mathcal{N}}^H ,
\label{eq:flex_covariance_decomposition}
\end{align}
where both terms are positive semidefinite. The first term can vanish only if $\mathbf{P}_{\mathcal{N}}\mathbf{J}_j=-\mathbf{I}$ over the entire perturbation subspace spanned by the support of $\boldsymbol{\Sigma}_{\bar{\Omega}}$. For a nontrivial perturbation subspace, this identity constrains the predictor Jacobian to a proper affine subset of the admissible Jacobian space, which has Lebesgue measure zero \cite{rudin1987real}. Consequently, the first term is nonzero for almost every admissible predictor Jacobian. Moreover, perturbations of acquired coefficients provide an additional positive-semidefinite contribution whenever the predictor maps them onto unacquired coefficients. Hence, for $0<\gamma_j<1$, the FLEX-MRI residual $\hat{\mathbf{r}}_j^{(z)}$ has nonzero conditional covariance after data consistency for almost every admissible predictor Jacobian. \textit{FLEX-MRI therefore realizes a non-degenerate residual term that is induced through the same form of randomized k-space support change as the target finite-transition variation, enters the same mean-plus-residual decomposition, and occupies the same measurement-null subspace as $\mathbf{r}_{t_{j-1}\leftarrow t_j}$.}

\begin{figure*}[t]
    \centering
    \includegraphics[width=\textwidth]{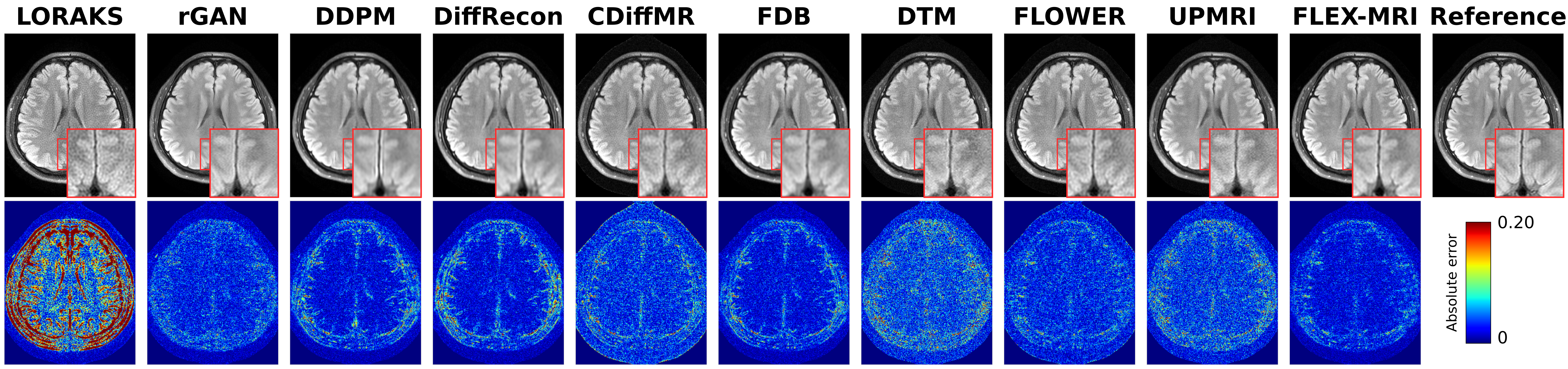}
    \caption{Representative reconstructions on fastMRI for a FLAIR-weighted cross-section at $R = 8$, shown alongside the fully-sampled reference. A zoomed region of interest is inset in each reconstruction, and the corresponding error map with respect to the reference is displayed beneath.}
    \label{fig:results_fastmri}
\end{figure*}

\section{Methods}

\subsection{Datasets}
Experiments were performed on multi-coil complex k-space data from the high-field fastMRI \cite{zbontar2018fastmri} and low-field M4Raw \cite{lyu2023m4raw} datasets, covering T\textsubscript{1}, T\textsubscript{2}, and FLAIR contrasts. Subjects were split into non-overlapping training, validation, and test sets. In M4Raw, a split of (128, 30, 25) subjects was used, yielding a total of 9,000 cross-sections. In fastMRI, a split of (240, 60, 120) was used with a total of 12,600 cross-sections; subjects whose volumes contained at least 10 cross-sections and more than 5 coil elements were analyzed. Geometric coil compression was applied to fastMRI to derive 5 virtual coils, preserving over 90\% of the data variance \cite{zhang2013coil}; M4Raw's native 4-channel acquisition was used directly. Two-dimensional retrospective undersampling at rates $R =4-8$ was applied with a normal density distribution \cite{lustig2007sparse}. To construct the imaging operators, coil sensitivities were estimated using ESPIRiT on a central calibration region \cite{uecker2014espirit}. Volumetric $k$-space data were inverse Fourier transformed and split across the readout dimension, and each cross-section was reconstructed individually.

\subsection{Competing Methods}
FLEX-MRI was compared against nine methods spanning traditional low-rank reconstruction (LORAKS \cite{haldar2013low}), adversarial reconstruction (rGAN \cite{dar2020prior}), Gaussian diffusion (DDPM \cite{song2022}, $T=1000$; DiffRecon \cite{peng2022towards}, $T=1000$), non-Gaussian diffusion (CDiffMR \cite{huang2023cdiffmr}, $T=1000$; FDB \cite{mirza2026learning}, $T=1000$), flow matching (FLOWER \cite{pourya2026flower}, $T=10$; UPMRI \cite{luo2025upmri,luo2025unsupervised}, $T=10$), and transition matching (DTM \cite{shaul2025transition}, $T=50$). For LORAKS, the $k$-space neighborhood radius and matrix rank were set to $(2,6)$ for M4Raw and $(2,30)$ for fastMRI; for rGAN, adversarial and pixel-wise loss weights were set to $1$ and $100$, respectively. For FLEX-MRI, the finite-transition predictor $\mu_\theta$ was implemented with a U-Net backbone \cite{ronneberger2015unet}. The endpoint state $\mathbf{x}_T$ was defined at $R=6$, recovery used $J=10$ retained transitions, and the current state was perturbed before each transition using a constant rate of $\gamma_j=0.1$.

\begin{table}[t]
    \centering \captionsetup{justification=justified,singlelinecheck=false}
    \caption{Reconstruction performance on fastMRI at R = 4-8. PSNR (dB) and SSIM (\%) are listed as mean$\pm$std. across the test set. Boldface denotes the highest performing method.}
    \resizebox{\columnwidth}{!}{%
    \begin{tabular}{|l|c|c|c|c|c|c|}
        \cline{2-7}
        \multicolumn{1}{c|}{} & \multicolumn{2}{c|}{$R = 4$} & \multicolumn{2}{c|}{$R = 6$} & \multicolumn{2}{c|}{$R = 8$} \\
        \cline{2-7}
        \multicolumn{1}{c|}{} & PSNR & SSIM & PSNR & SSIM & PSNR & SSIM \\
        \hline
        LORAKS & 35.2$\pm$2.3 & 92.0$\pm$4.8 & 33.5$\pm$2.2 & 90.5$\pm$5.2 & 32.3$\pm$2.2 & 89.5$\pm$5.4 \\
        \hline
        rGAN & 35.9$\pm$2.8 & 93.1$\pm$4.7 & 33.9$\pm$2.6 & 90.7$\pm$5.5 & 32.9$\pm$2.8 & 90.1$\pm$5.9 \\
        \hline
        DDPM & 36.1$\pm$2.2 & 93.7$\pm$4.0 & 34.8$\pm$2.3 & 92.2$\pm$5.2 & 34.0$\pm$2.2 & 91.1$\pm$4.9 \\
        \hline
        DiffRecon & 35.7$\pm$2.3 & 93.6$\pm$4.5 & 34.5$\pm$2.2 & 92.1$\pm$5.1 & 33.7$\pm$2.2 & 91.0$\pm$5.5 \\
        \hline
        CDiffMR & 31.4$\pm$2.7 & 78.5$\pm$8.1 & 31.2$\pm$2.4 & 79.1$\pm$7.2 & 31.2$\pm$2.3 & 80.5$\pm$7.4 \\
        \hline
        FDB & 37.4$\pm$2.3 & 94.3$\pm$3.9 & 35.9$\pm$2.4 & 92.7$\pm$5.2 & 35.0$\pm$2.2 & 91.5$\pm$4.9 \\
        \hline
        DTM & 30.2$\pm$1.4 & 75.9$\pm$4.9 & 29.2$\pm$1.3 & 72.9$\pm$5.3 & 28.7$\pm$1.3 & 71.2$\pm$5.5 \\
        \hline
        FLOWER & 33.8$\pm$2.5 & 84.5$\pm$6.1 & 33.4$\pm$2.3 & 84.6$\pm$6.0 & 33.1$\pm$2.2 & 84.9$\pm$6.0 \\
        \hline
        UPMRI & 36.6$\pm$2.1 & 93.8$\pm$4.3 & 34.2$\pm$2.4 & 89.7$\pm$7.7 & 31.9$\pm$2.9 & 82.7$\pm$11.7 \\
        \hline
        FLEX-MRI & \textbf{38.2$\pm$2.5} & \textbf{94.5$\pm$4.3} & \textbf{36.7$\pm$2.3} & \textbf{93.5$\pm$4.7} & \textbf{35.8$\pm$2.2} & \textbf{92.7$\pm$5.0} \\
        \hline
    \end{tabular}
    }
    \label{tab:fastmri}
\end{table}

Learned models represented real and imaginary components with separate channels and were trained for 50 epochs on M4Raw and 100 epochs on fastMRI using Adam with a learning rate of $\eta=10^{-4}$ and $(\beta_1,\beta_2)=(0.5,0.9)$. Method-specific hyperparameters were selected by cross-validation, and all methods used identical data splits and undersampling patterns. Significance of differences between methods was assessed via a non-parametric Wilcoxon signed-rank test.

\begin{figure*}[t]
    \centering
    \includegraphics[width=\textwidth]{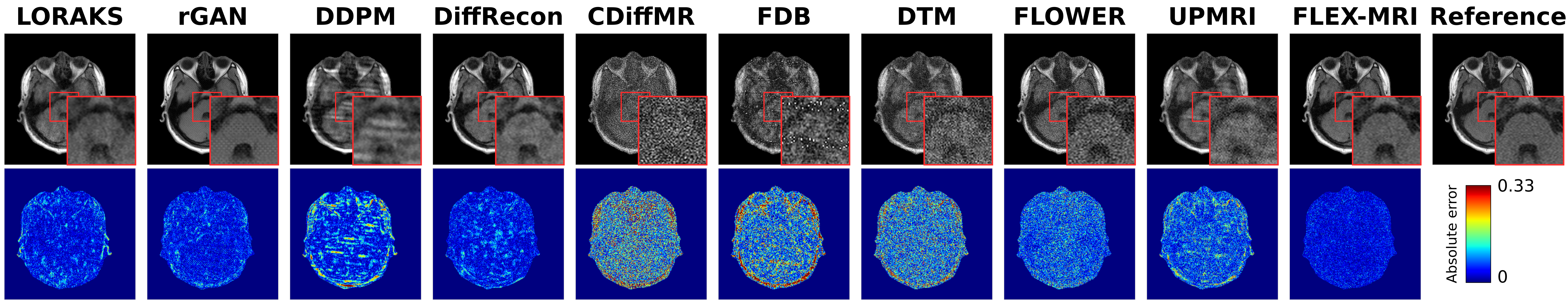}
    \caption{Representative reconstructions on M4Raw for a T\textsubscript{1}-weighted cross-section at $R = 4$, shown alongside the fully-sampled reference. A zoomed region of interest is inset in each reconstruction, and the corresponding error map with respect to the reference is displayed beneath.}
    \label{fig:results_m4raw}
\end{figure*}

\section{Results}

\subsection{Comparison Studies}
Table~\ref{tab:fastmri} and Fig.~\ref{fig:results_fastmri} summarize results on the higher-field fastMRI dataset. FLEX-MRI attains the highest PSNR and SSIM across $R$ ($p<0.05$). Averaged across $R$ and methods within each family, FLEX-MRI improves over the diffusion baselines by 2.7dB PSNR and 4.4\% SSIM, and over the few-step flow- and transition-matching baselines by 4.6dB and 11.3\%. From $R=4$ to $8$, average PSNR and SSIM decrease by 1.7dB and 1.5\% for diffusion methods, compared with 2.3dB and 5.1\% for few-step baselines. In comparison, FLEX-MRI shows decreases of 2.4dB and 1.8\%, maintaining greater SSIM stability than the few-step baselines and behavior closer to the diffusion family in this respect. The representative reconstructions in Fig.~\ref{fig:results_fastmri} reflect these trends. DDPM, DiffRecon and FDB suppress residual noise effectively but show greater edge errors consistent with smoothing of fine anatomical boundaries, whereas rGAN, CDiffMR and more prominently the flow- and transition-matching baselines yield high-frequency residual structure consistent with noise amplification. FLEX-MRI effectively suppresses these residual errors while preserving sharper anatomical boundaries.

Table~\ref{tab:m4raw} and Fig.~\ref{fig:results_m4raw} show the corresponding results on the low-field M4Raw dataset, where noise is more prominent. FLEX-MRI again attains the highest PSNR and SSIM across $R$ ($p<0.05$). On average, FLEX-MRI improves over the diffusion baselines by 4.1dB PSNR and 35.1\% SSIM, and over the few-step baselines by 2.1dB and 23.3\%, respectively. The relative behavior of the competing families changes substantially from fastMRI: flow-based methods become more competitive in PSNR, while DiffRecon remains more robust than the other diffusion baselines under the noisier acquisition conditions. From $R=4$ to $8$, average PSNR and SSIM decrease by 0.7dB and 2.2\% for diffusion baselines and by 1.5dB and 8.4\% for few-step baselines, whereas FLEX-MRI decreases by 0.7dB and 2.6\%. The qualitative results show a corresponding change in failure modes. DDPM and DiffuseRecon exhibit structured residual striping, whereas CDiffMR and FDB show more pronounced pixelation artifacts. DTM and the flow-based methods suppress these artifacts more effectively than the latter diffusion baselines, although visible high-frequency residual structure remains, while rGAN is comparatively robust in this example but retains localized errors. FLEX-MRI more effectively suppresses these residual patterns while preserving the underlying anatomical structure.

Finally, Table~\ref{tab:cross-dataset} reports cross-dataset generalization, where models trained on M4Raw are tested on fastMRI. FLEX-MRI attains the highest PSNR and SSIM across $R$ ($p<0.05$). On average, FLEX-MRI improves over the diffusion baselines by 7.0dB PSNR and 12.9\% SSIM, and over the few-step baselines by 3.2dB and 5.3\%. Notably, cross-dataset transfer reduces FLEX-MRI performance by only 0.8dB PSNR and 0.4\% SSIM on average relative to in-domain training on fastMRI (Table~\ref{tab:fastmri}). These results show that FLEX-MRI generalizes effectively across substantial differences in field strength and acquisition characteristics without retraining.

\subsection{Computational Efficiency}
Table~\ref{tab:efficiency} compares inference time and peak GPU memory across methods. The diffusion-based methods operating over finely discretized trajectories require substantially longer reconstruction times, whereas FLEX-MRI operates in the same sub-second regime as the few-step transition- and flow-matching methods. FLEX-MRI is approximately two orders of magnitude faster than the 1000-step diffusion baselines while remaining comparable in inference time to DTM, FLOWER and UPMRI. Peak memory follows a similar practical trend: FLEX-MRI requires 1.05~GB, matching the few-step baselines and remaining below the 1.71--3.19~GB required by the diffusion methods. Thus, the reconstruction gains in Tables~\ref{tab:fastmri} and \ref{tab:m4raw} are obtained while retaining the computational efficiency and moderate memory requirements of few-step recovery.

\begin{table}[t]
    \centering \captionsetup{justification=justified,singlelinecheck=false}
    \caption{Reconstruction performance on M4Raw at R = 4-8.}
    \resizebox{\columnwidth}{!}{%
    \begin{tabular}{|l|c|c|c|c|c|c|}
        \cline{2-7}
        \multicolumn{1}{c|}{} & \multicolumn{2}{c|}{$R = 4$} & \multicolumn{2}{c|}{$R = 6$} & \multicolumn{2}{c|}{$R = 8$} \\
        \cline{2-7}
        \multicolumn{1}{c|}{} & PSNR & SSIM & PSNR & SSIM & PSNR & SSIM \\
        \hline
        LORAKS & 22.4$\pm$1.6 & 76.3$\pm$2.7 & 22.2$\pm$1.6 & 74.2$\pm$2.0 & 22.0$\pm$1.6 & 72.9$\pm$3.3 \\
        \hline
        rGAN & 22.5$\pm$1.6 & 68.1$\pm$2.5 & 22.4$\pm$1.6 & 67.1$\pm$2.7 & 22.3$\pm$1.5 & 66.5$\pm$2.7 \\
        \hline
        DDPM & 22.1$\pm$1.2 & 41.2$\pm$4.7 & 21.7$\pm$1.1 & 38.5$\pm$4.6 & 21.5$\pm$1.1 & 36.9$\pm$4.5 \\
        \hline
        DiffRecon & 22.9$\pm$1.3 & 46.1$\pm$5.0 & 22.7$\pm$1.3 & 43.8$\pm$4.8 & 22.5$\pm$1.3 & 42.4$\pm$4.7 \\
        \hline
        CDiffMR & 16.7$\pm$1.2 & 34.6$\pm$2.1 & 16.7$\pm$1.2 & 34.5$\pm$2.4 & 16.5$\pm$1.1 & 34.0$\pm$2.2 \\
        \hline
        FDB & 17.7$\pm$1.1 & 44.3$\pm$5.1 & 16.6$\pm$1.1 & 44.1$\pm$4.9 & 16.0$\pm$1.1 & 44.0$\pm$4.8 \\
        \hline
        DTM & 20.4$\pm$1.1 & 48.3$\pm$4.7 & 19.8$\pm$1.1 & 45.4$\pm$4.9 & 19.5$\pm$1.1 & 44.1$\pm$4.9 \\
        \hline
        FLOWER & 23.1$\pm$1.4 & 55.3$\pm$3.5 & 22.7$\pm$1.4 & 54.2$\pm$3.4 & 22.5$\pm$1.3 & 54.2$\pm$3.5 \\
        \hline
        UPMRI & 23.3$\pm$1.6 & 67.1$\pm$9.8 & 21.3$\pm$1.7 & 53.4$\pm$10.6 & 20.2$\pm$1.6 & 47.3$\pm$9.9 \\
        \hline
        FLEX-MRI & \textbf{23.9$\pm$1.6} & \textbf{76.9$\pm$2.5} & \textbf{23.5$\pm$1.5} & \textbf{75.1$\pm$2.8} & \textbf{23.2$\pm$1.5} & \textbf{74.3$\pm$3.0} \\
        \hline
    \end{tabular}
    }
    \label{tab:m4raw}
\end{table}

\begin{table}[t]
    \centering \captionsetup{justification=justified,singlelinecheck=false}
    \caption{Cross-dataset generalization at $R = 4$--$8$. Models were trained on M4Raw and tested on fastMRI.}
    \resizebox{\columnwidth}{!}{%
    \begin{tabular}{|l|c|c|c|c|c|c|}
        \cline{2-7}
        \multicolumn{1}{c|}{} & \multicolumn{2}{c|}{$R = 4$} & \multicolumn{2}{c|}{$R = 6$} & \multicolumn{2}{c|}{$R = 8$} \\
        \cline{2-7}
        \multicolumn{1}{c|}{} & PSNR & SSIM & PSNR & SSIM & PSNR & SSIM \\
        \hline
        LORAKS & 35.2$\pm$2.3 & 92.0$\pm$4.8 & 33.5$\pm$2.2 & 90.5$\pm$5.2 & 32.3$\pm$2.2 & 89.5$\pm$5.4 \\
        \hline
        rGAN & 33.5$\pm$2.1 & 90.5$\pm$2.9 & 31.6$\pm$2.2 & 87.8$\pm$3.5 & 30.4$\pm$2.3 & 85.9$\pm$3.9 \\
        \hline
        DDPM & 35.0$\pm$2.3 & 89.6$\pm$4.9 & 33.1$\pm$2.2 & 86.0$\pm$5.8 & 31.9$\pm$2.2 & 83.3$\pm$6.2 \\
        \hline
        DiffRecon & 35.5$\pm$2.0 & 93.1$\pm$3.4 & 33.3$\pm$2.1 & 90.9$\pm$4.0 & 31.9$\pm$2.1 & 89.3$\pm$4.3 \\
        \hline
        CDiffMR & 35.0$\pm$1.7 & 91.5$\pm$3.3 & 33.0$\pm$1.6 & 87.6$\pm$3.9 & 31.4$\pm$1.5 & 83.9$\pm$4.3 \\
        \hline
        FDB & 18.0$\pm$5.6 & 56.7$\pm$14.6 & 15.6$\pm$3.1 & 54.6$\pm$9.9 & 15.0$\pm$2.5 & 56.2$\pm$8.6 \\
        \hline
        DTM & 33.2$\pm$1.9 & 87.3$\pm$3.9 & 31.6$\pm$1.8 & 84.3$\pm$4.5 & 30.6$\pm$1.7 & 82.2$\pm$4.9 \\
        \hline
        FLOWER & 35.7$\pm$2.2 & 91.0$\pm$4.1 & 34.1$\pm$2.1 & 88.8$\pm$4.6 & 33.1$\pm$1.9 & 87.3$\pm$4.9 \\
        \hline
        UPMRI & 34.5$\pm$1.7 & 92.8$\pm$3.2 & 32.3$\pm$1.7 & 90.2$\pm$3.8 & 30.7$\pm$1.7 & 87.2$\pm$4.6 \\
        \hline
        FLEX-MRI & \textbf{37.6$\pm$2.3} & \textbf{94.2$\pm$3.3} & \textbf{35.9$\pm$2.1} & \textbf{93.1$\pm$3.7} & \textbf{34.8$\pm$1.9} & \textbf{92.2$\pm$3.9} \\
        \hline
    \end{tabular}
    }
    \label{tab:cross-dataset}
\end{table}

\begin{table*}[t]
    \centering
    \begin{minipage}[t]{0.21\textwidth}
        \vspace{0pt}
        \captionsetup{justification=justified,singlelinecheck=false}
        \caption{Inference time (s) and memory use (GB) for a single cross-section on an Nvidia RTX 4090.}
        \label{tab:efficiency}
    \end{minipage}
    \hfill
    \begin{minipage}[t]{0.74\textwidth}
        \vspace{0pt}
        \centering
        \setlength{\tabcolsep}{3.0pt}
        \renewcommand{\arraystretch}{1.20}
        \resizebox{\linewidth}{!}{%
        \begin{tabular}{|l|c|c|c|c|c|c|c|c|c|c|}
            \hline
            & LORAKS & rGAN & DDPM & DiffRec. & CDiffMR & FDB & DTM & FLOWER & UPMRI & FLEX-MRI \\
            \hline
            Time (s) & 0.180 & 0.0056 & 37.12 & 36.26 & 34.03 & 59.09 & 0.437 & 0.477 & 0.318 & 0.349 \\
            \hline
            Memory (GB) & 0.05 & 0.13 & 1.72 & 1.71 & 2.23 & 3.19 & 1.05 & 1.05 & 1.05 & 1.05 \\
            \hline
        \end{tabular}%
        }
    \end{minipage}
\end{table*}

\subsection{Ablation Studies}
\label{subsec:ablation}
Ablation studies on fastMRI examine the principal design elements of FLEX-MRI: support-relaxed finite-transition learning and inference-time stochastic exploration. Table~\ref{tab:ablation-perturbation} compares the proposed configuration with controlled variants that alter the training input, inference-time perturbation, or finite-transition objective. Replacing stochastic exploration with deterministic inference consistently reduces reconstruction quality, while replacing support-relaxed learning with the support-restricted objective also degrades performance. Perturbing the training inputs provides no additional benefit. These results support the proposed combination of forward-process states during training, support-relaxed finite-transition learning, and stochastic perturbation confined to inference.

\begin{figure}[t]
    \centering
    \includegraphics[width=\columnwidth]{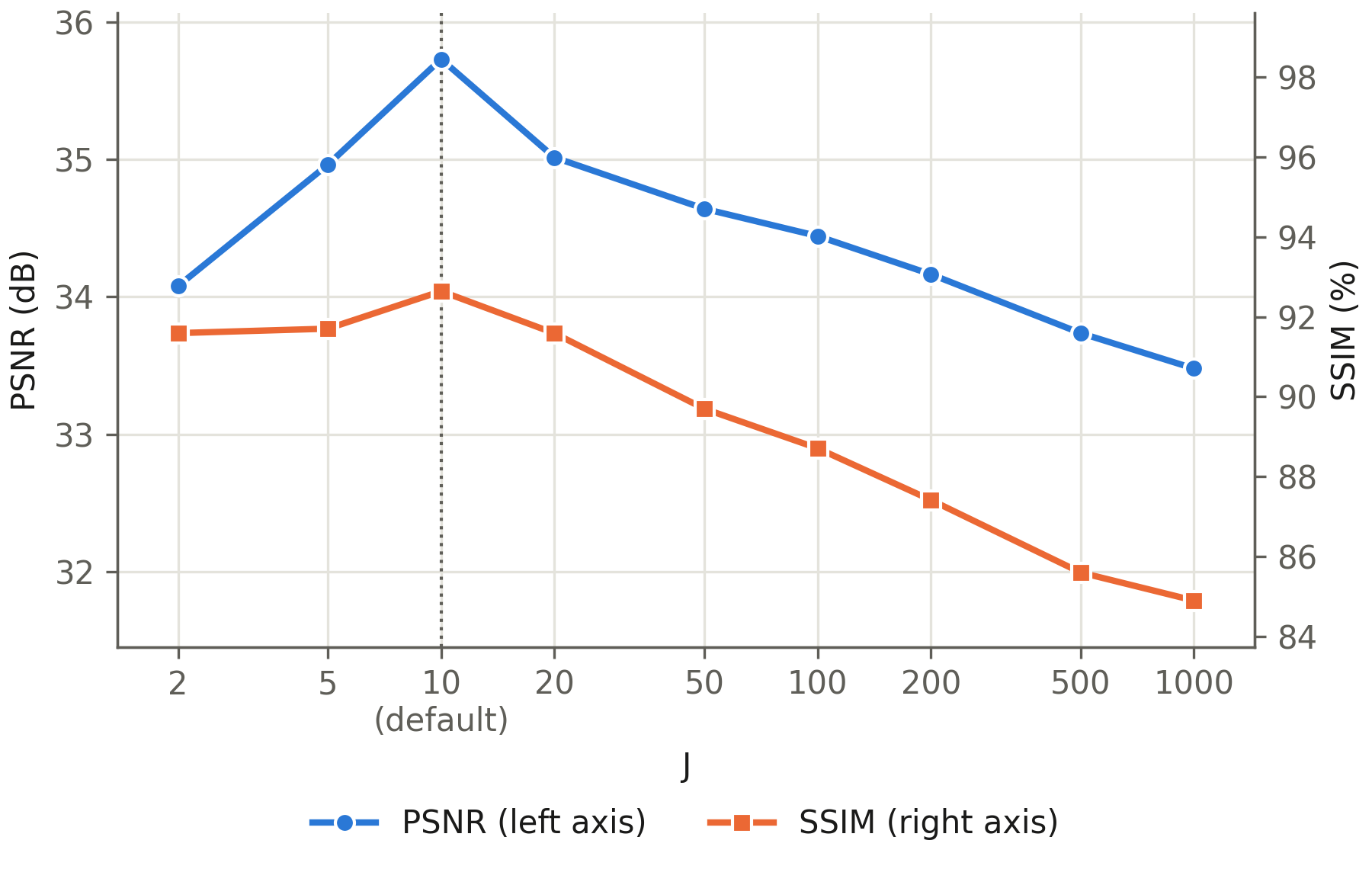}
    \caption{Effect of the number of retained transitions $J$ on reconstruction quality on fastMRI at $R = 8$. Mean PSNR (dB, left axis) and SSIM (\%, right axis) are plotted for $J$ ranging from 2 to 1000 on a logarithmic scale. The dotted line marks the adopted $J = 10$.}
    \label{fig:J_ablation}
\end{figure}

Fig.~\ref{fig:J_ablation} further examines the number of retained transitions $J$, which determines the spacing between consecutive retained states along the recovery trajectory. Reconstruction quality peaks at the adopted setting of $J=10$ and decreases for both smaller and larger values. With fewer transitions, each retained transition spans a larger portion of the trajectory, requiring a single predictor evaluation to account for a larger mean displacement and greater residual variation, as described in Sec.~\ref{subsec:finite-transition-theory}. At $J=2$, PSNR and SSIM are 1.7~dB and 1.1\% lower than at $J=10$, respectively. For $J>10$, reconstruction quality decreases monotonically up to $J=1000$, where PSNR and SSIM are 2.3~dB and 7.8\% lower than at $J=10$. Because the perturbation rate is held fixed per transition, increasing $J$ introduces progressively more perturbation events while reducing the recovery update associated with each individual transition. The observed decline at large $J$ is therefore consistent with an unfavorable balance between repeated stochastic exploration and increasingly small finite-transition updates, together with accumulation of prediction errors across a longer sequence of transitions. Since inference cost also scales approximately linearly with $J$, the intermediate setting of $J=10$ provides the highest reconstruction quality while retaining low computational cost.

\begin{table}[t]
    \centering
    \captionsetup{justification=justified,singlelinecheck=false}
    \caption{Ablation of the inference mode, transition support, and training input.
    PSNR (dB) and SSIM (\%) are reported on the fastMRI test set at $R=8$.}
    \setlength{\tabcolsep}{5pt}
    \resizebox{0.85\columnwidth}{!}{%
    \begin{tabular}{|c|c|c|c|c|}
        \hline
        \makecell{Inference\\mode} &
        \makecell{Transition\\support} &
        \makecell{Training\\input} &
        PSNR & SSIM \\
        \hline
        \multirow{4}{*}{Stochastic}
        & \multirow{2}{*}{Relaxed}
        & Nominal
        & \textbf{35.8$\pm$2.2}
        & \textbf{92.7$\pm$5.0} \\
        \arrayrulecolor{gray!35}\cline{3-5}\arrayrulecolor{black}
        &
        & Perturbed
        & 35.4$\pm$2.2
        & 92.2$\pm$4.9 \\
        \arrayrulecolor{gray!35}\cline{2-5}\arrayrulecolor{black}
        &
        \multirow{2}{*}{Restricted}
        & Nominal
        & 35.0$\pm$2.1
        & 91.5$\pm$5.1 \\
        \arrayrulecolor{gray!35}\cline{3-5}\arrayrulecolor{black}
        &
        & Perturbed
        & 34.9$\pm$2.1
        & 91.6$\pm$5.1 \\
        \hline
        \multirow{4}{*}{Deterministic}
        & \multirow{2}{*}{Relaxed}
        & Nominal
        & 34.9$\pm$2.3
        & 91.1$\pm$5.2 \\
        \arrayrulecolor{gray!35}\cline{3-5}\arrayrulecolor{black}
        &
        & Perturbed
        & 34.5$\pm$2.2
        & 91.0$\pm$5.0 \\
        \arrayrulecolor{gray!35}\cline{2-5}\arrayrulecolor{black}
        &
        \multirow{2}{*}{Restricted}
        & Nominal
        & 34.2$\pm$2.3
        & 89.9$\pm$5.1 \\
        \arrayrulecolor{gray!35}\cline{3-5}\arrayrulecolor{black}
        &
        & Perturbed
        & 34.0$\pm$2.3
        & 90.0$\pm$5.3 \\
        \hline
    \end{tabular}}
    \label{tab:ablation-perturbation}
\end{table}

\section{Discussion}
Our results show that FLEX-MRI consistently improves reconstruction quality while retaining the computational efficiency of few-step recovery. The theoretical analysis links stochastic finite-transition recovery to residual variation induced by k-space perturbation, which remains within the degrees of freedom unresolved by the acquired measurements. The ablation studies further support this interpretation: reconstruction quality decreases when stochastic exploration is removed or support-relaxed learning is replaced with support-restricted learning, whereas perturbing the training inputs provides no further improvement.

One natural extension of FLEX-MRI is to adapt stochastic exploration along the recovery trajectory rather than using a fixed perturbation rate and number of retained transitions. The amount and structure of unresolved information change as recovery progresses, suggesting that transition-dependent perturbation rates or spacing between retained states could provide a more selective allocation of stochastic exploration. Related reconstruction studies have explored adaptive generative priors and multi-scale physics-driven recovery \cite{gungor2022adaptive,mambaroll,hammernik2022physicsdriven}. Extending the present covariance analysis beyond local linearization could similarly provide a more complete characterization of stochastic propagation through strongly nonlinear finite-transition predictors.

The current experiments focus on supervised model training for reconstructing two-dimensional Cartesian MR acquisitions. Evaluating non-Cartesian sampling, volumetric reconstruction, and broader scanner and protocol variation would help establish how finite-transition recovery generalizes across acquisition settings \cite{fedgat}. Recent work has examined multi-institutional learning as a means of broadening the training distribution, while self-supervised reconstruction provides an alternative when fully-sampled reference data are unavailable \cite{korkmaz2023selfsupervised,cukurtutorial2026,yaman2022multimask,wang2024review}. Incorporating such strategies into finite-transition learning could therefore extend FLEX-MRI to settings where either training distributions or reference data are more limited.

Because FLEX-MRI generates alternative recovery trajectories from the same acquired measurements, repeated reconstructions could also be used to characterize spatial uncertainty in the recovered image. Existing work has quantified reconstruction uncertainty through posterior sampling or propagation of measurement noise through learned reconstruction models \cite{luo2022mri,dalmaz2025noise}. An important question is whether variation across FLEX-MRI trajectories provides complementary information about image content that remains weakly constrained by the measurements. Such analyses should ultimately be accompanied by radiological evaluation, since reference-image distortion metrics alone cannot establish whether stochastic reconstruction preserves diagnostically relevant structures or introduces unsupported image content \cite{bhadra2021hallucinations,cukurtutorial2026}.

\section{Conclusion}
Here we presented FLEX-MRI, a physics-guided finite-transition method for accelerated MRI reconstruction that combines progressive k-space undersampling, support-relaxed learning, and inference-time stochastic exploration without an auxiliary generative sampler. We showed theoretically that stochastic k-space perturbation induces a non-degenerate transition-level residual within the degrees of freedom left unresolved by the acquired measurements and retains nonzero variation after propagation through the predictor and data-consistency projections. Across two multi-coil MRI datasets, FLEX-MRI consistently improved reconstruction quality over competing generative methods while operating roughly two orders of magnitude faster than diffusion-based baselines.

\bibliographystyle{IEEEtran}
\bibliography{IEEEabrv,refs}

\end{document}